\documentstyle[psfig,graphics]{mn}

\begin{document}

\title{Temporal properties of short GRBs}

\author[Ehud Nakar \& Tsvi Piran]
{Ehud Nakar \& Tsvi Piran\\
  Racah Institute, Hebrew University, Jerusalem 91904, Israel}

\maketitle
\begin{abstract}
{We analyze a sample of bright short bursts from the BATSE
4B-catalog and find that many short bursts are highly variable
(\( \delta t_{min}/T\ll 1 \), where \( \delta t_{min} \) is the
shortest pulse duration and \( T \) is the burst duration). This
indicates that it is unlikely that short bursts are produced by
external shocks. We also  analyze the available (first 1-2
seconds)  high resolution (TTE) data of some of the long bursts.
We find that variability on a 10ms time-scale is common in long
bursts. This result shows that some long bursts are even more
variable than it was thought before (\( \delta t_{min}/T\approx
10^{-4}-10^{-3} \)).}
\end{abstract}


\section{Introduction}

The temporal features of GRBs are among the more interesting
clues on their origin.  The temporal features of long bursts (\(
T_{90}>2sec \)) were widely investigated (e.g. Norris 1995; Norris
et al. 1996; Lee, Bloom \& Scargle 1995; Beloborodov Stern \& Svensson,
2000), while only a few works (Scargle, Norris \& Bonnel 1997;
Cline, Matthey \& Otwinowski 1999) discuss the temporal structure
of short (\( T_{90}<2sec \)) bursts. This is not surprising. Short
bursts are much harder to analyze because of their significantly
lower signal to noise ratios.  We developed a new algorithm that
is sensitive enough to identify pulses in short bursts. Using
this algorithm we search here for subpulses in short bursts and
determine their duration (\( \delta t \)). This enables us to set
upper limits on the shortest time scale seen in the bursts  to
set limits on the variability of short bursts.

Sari \& Piran (1997) and  Fenimore, Madras, \& Nayakshin (1996)
have shown that angular spreading would smooth any variability
produced by external shocks (unless the GRB production is very
inefficient with the efficiency of the order of \( (\delta
t_{min}/T) \) defined below (Sari \& Piran, 1997)).   The
critical parameter in this analysis is the ratio between the
shortest time scale, on which the burst varies significantly, and
the longest time scale in the burst. This motivates us to focus
here on  the ratio \( \delta t_{min}/T \), where $\delta t_{min}$
is the minimal observed duration of an individual pulse and $T$ is
the duration of the burst.  High variability means low \( \delta
t_{min}/T \) (\( \ll 1 \)) values while for smooth bursts \(
\delta t_{min}/T \approx 1 \). Our aim is to explore whether
short bursts can be produced by external shocks. However, our
analysis is not directed by this motivation and the results
concerning the variability of short bursts are valid
independently of this motivation.

In the second part of this paper we compare the shortest time
scales of long and short bursts. We analyze (using the same
algorithm) the high resolution (TTE) data of long bursts.
Unfortunately, this data is available only for the first 1-2
seconds of each burst, so we can analyze only a small fraction of
each long burst. still we are able to demonstrate that very short
time scales (10ms or less) are common.

Our  analysis deals with statistically significant individual
pulses. We define an observed peak in the light curve as the
highest count rate within a series of counts that is
statistically significant (more than 4 $\sigma$) above the counts
at some time before and some time   after it. Each peak
corresponds to a pulse. The width of the pulse is determined by
the width at a quarter of the maximum, or, rarely,  by the minima
between neighboring peaks if the counts do not drop below quarter
of the maximum (see the Appendix for a detailed definition and
for a discussion of the algorithm). An elementary pulse in the
observed light curve must not necessarily correspond directly to
an elementary emission event in the source.  A single pulse in the
light curve could, in principle, be  composed of numerous
emission events. In this case the emission process is even more
variable then the observed light curve and the results we obtain
here should be considered only as upper limits to the intrinsic
variability of the sources.

We find that most short bursts (although not all of them) are
highly variable (\( \delta t_{min}/T\ll 1 \)) \textbf{.} When
analyzing the high resolution data of long burst  we find that the
shortest time scales seen in long bursts are similar to those in
short bursts. This result is limited to the highest resolution in
which we analyzed the long bursts - 5ms. Such time scales were
already observed in at least one long burst (Lee, Bloom \&
Petrosian 2000). We show here that these time scales are common.

In section 2 we describe the data samples considered in our
analysis. We describe the results in section 3 and we discuss
their implications  in section 4. Our algorithm is described in
the appendix.

\section{The data samples}

We examine bright short and long bursts from BATSE 4B-catalog. We
use two BATSE data formats: the 64ms concatenate data and the TTE
data (see Scargle (1998) for a detailed review). The 64ms
concatenate data includes the photon counts of each burst, in a
64ms time bins, from a few seconds before the burst trigger till
a few hundred of seconds after the trigger. The concatenate data
includes also very early and very late data of the burst in a
1024ms resolution. We use only the 64ms resolution data. The TTE
(Time Tagged Events) data includes the arrival time of each
photon in a \( 2\mu sec \) resolution. This data contains only
records of the first 1-2 seconds of each burst. Hence it contains
whole short bursts, but only  a fraction of long bursts. Both
data formats have four energy channels. We use the sum of all the
channels (in both formats), that is photon energy E \( > \) 25Kev.

We consider several samples. We consider a sample of short bursts
(denoted \textit{`short'}) and a comparable sample of long bursts
(denoted \textit{'long'}). However, the properties of the
\textit{'long'} sample cannot be compared directly with those of
the \textit{'short'} sample. The long bursts data is binned in
longer time bins then the short bursts data. Therefore two equally
intense bursts (one short and one long) would have a different
signal-to-noise ratio (S/N). Hence, we generate a third sample
denoted \textit{'noisy long'} by adding noise to the
\textit{'long'} sample so that the S/N  of the bursts in this sample
would be comparable to the S/N of the \textit{'short'} sample.
Finally, in order to determine the shortest time scale in long
bursts, we consider a sample of long bursts with a good TTE
coverage of the first second. This sample is called \textit{'high
res long'}.

\subsection{The \textit{'short'} data sample\label{shorts}}

There are about 400 records of short bursts in the BATSE
4B-catalog. However most of these bursts are too faint and it is
impossible to retrieve their temporal features. Furthermore, not
all short bursts have a good TTE coverage. There is a trade-off
between the sample size, the resolution and the signal to noise
ratio. We consider, here,  a sample of the brightest 33 short
bursts (peak flux in 64ms\( > \)4.37\( ph/(sec\cdot cm^{2}) \))
with a good TTE data coverage. In order to get a reasonable
signal to noise ratio we have binned this data into 2ms time
bins. In this resolution the S/N of the brightest peak in the
faintest burst (from our sample) is 4.7. As described in the
appendix we consider a peak as statically significant only if it
is more then 4\( \sigma  \) above the background. Hence this is
the largest sample we could consider. The minimal recognized
pulse width with this resolution is 4ms.

\subsection{The \textit{'long'} data samples\label{noisy}}

We need a sample of long bursts that could be compared to the \textit{'short'}
sample. \label{longs}The first sample we considered is a sample of 34 long
bursts (called \textit{'long'} sample) with the same 64ms peak fluxes (one to
one) as the bursts in the \textit{'short'} sample. This sample contain fairly
bright long bursts, but not the brightest long bursts. This way we prevent differences
that arise from different brightness.

However, the  \textit{'long'} sample cannot be compared directly
with the \textit{'short'} sample. The  \textit{'long'} sample is
binned in 64ms bins while the \textit{'short'} sample is binned
into 2ms time bins. Therefore, assuming the same background noise
level, the S/N of the \textit{'long'} sample is larger by a factor
of \( \sqrt{32} \) then the S/N of the \textit{'short'} sample. To
obtain a comparable sample we produced another data sample
(denoted \textit{'noisy long'} sample). This data set is produced
by adding noise to the  \textit{'long'} sample. We treat this
sample as if the basic time bin is 2ms and add a Poisson noise
accordingly. For a given long burst with counts \( C(t) \) we
generate a  noisy signal, \( C_{noisy}(t) \), using the following
simple procedure. The noisy signal at time $t$, \( C_{noisy}(t)
\), is a Poisson variable (\(
P_{(C_{noisy}(t)=i)}=\frac{\lambda(t) ^{i}}{i!}e^{-\lambda(t) }
\)) with \( \lambda (t)\equiv C(t)/32 \). This noisy signal is
smaller by a factor of \( 64ms/2ms=32 \) than the original signal
and its standard deviation is correspondingly smaller by a factor 
of \( \sqrt{32} \)  then the original standard deviation. 
There is a minor caveat in this procedure.
The original signal contains its own noise, but since this
original noise is smaller by a factor of \( \sqrt{32} \) then the
added noise, it is negligible. The S/N ratio of the new
\textit{'noisy'} sample is comparable to the S/N ratio of the
\textit{'short'} sample.

We use the \textit{'noisy long'} sample to investigate the
influence of the noise on the analyzed temporal properties. We do
so by comparing the temporal properties of the \emph{'long'}
sample with the temporal properties of the \textit{'noisy long'}
\textit{\emph{sample.}} In this way we can  estimate what was the
original temporal structure of the \textit{'short'} sample.

\subsection{\label{longs start}High-resolution long bursts }

The comparison between the shortest time scales in long and short
bursts requires the analysis of high-resolution long bursts. The
only data with high enough resolution is the TTE data, which is
available only for the first 1-2sec of the bursts. We have
searched for long bursts that begin with a bright pulse during
the first two seconds, also demanding that the counts would drop
back to the background level during this time. This way the
beginning of the light curve is not be dominated by a pulse longer
then 2sec. We found 15 such bursts which we denoted as the
\textit{'high res long'} sample. We compared the first 1-2sec of
these bursts with 15 short bursts with comparable peak fluxes
(taken out of the \textit{'short'} sample). The analysis of both
groups is done in 5ms time bins. This sample is rather small and
not randomly chosen, but this is the best sample one could get
within the data limitation.

\section{Results}

\subsection{Attributes of long bursts pulses }

\begin{figure}

\psfig{figure=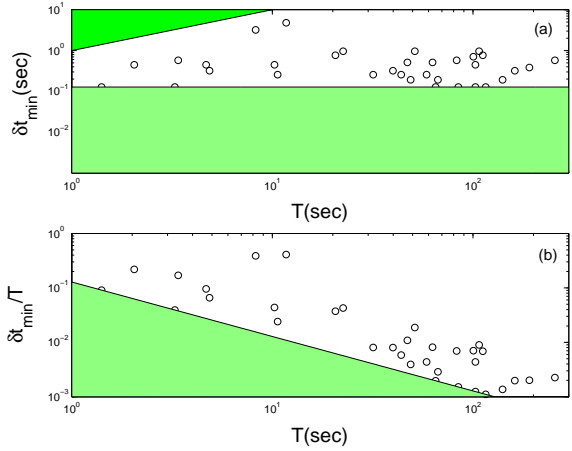,angle=0,height=0.25\textheight,width=0.9\columnwidth}

\caption{\label{longs dt/T fig} {\bf a)} \protect\( \delta
t_{min}\protect \) and {\bf b)} \protect\( \delta t_{min}/T\protect \)
as a function of the burst duration \protect\( T\protect \) in
long bursts. The gray areas are not allowed because of the
resolution (\protect\( \delta t>128ms\protect \)) or the
parameters definition \protect\( (\delta t \le T)\protect \) }
\end{figure}

We begin by estimating the duration, \( T \), and the shortest
pulse duration, \( \delta t_{min} \), of the \textit{`long'}
bursts. Fig. \ref{longs dt/T fig} shows \( \delta t_{min}/T \)
and \( \delta t_{min} \) as a function of \( T \). Fig.
\ref{longs dt/T fig}a shows that \( \delta t_{min} \) and \( T \)
are
not correlated\footnote{%
This implies, incidentally, that intrinsic effects and not
cosmological red-shifts dominate spread in \( T \) and \( \delta
t \). }. Consequently, \( \delta t_{min}/T \) is smaller for
longer bursts. The value of \( \delta t_{min}/T \) for the longer
bursts are \( 10^{-3}-10^{-2} \). The gray areas are restricted
because of the resolution.  \( \delta t_{min} \) is limited by
the resolution, suggesting that the bursts are variable even on
shorter time scales and therefore \( \delta t_{min}/T \) is even
smaller. This suggestion is confirmed later when we discuss the
high resolution data (see sec. \ref{long start result}).

\subsection{The effects of noise on the temporal structure\label{noisy vs longs}}

We turn now to the effect of noise on the time profile. We do so
by comparing the attributes of the \textit{'long'} sample with
these of the \textit{'noisy long'} sample (see
\ref{noisy}). This procedure also tests our algorithm.
Since we know the original signal (in the \textit{'long'} sample)
we can find out the efficiency of the algorithm in retrieving the
attributes of the \textit{'long'} sample out of the
\textit{'noisy long'} sample.

\begin{figure}
\psfig{figure=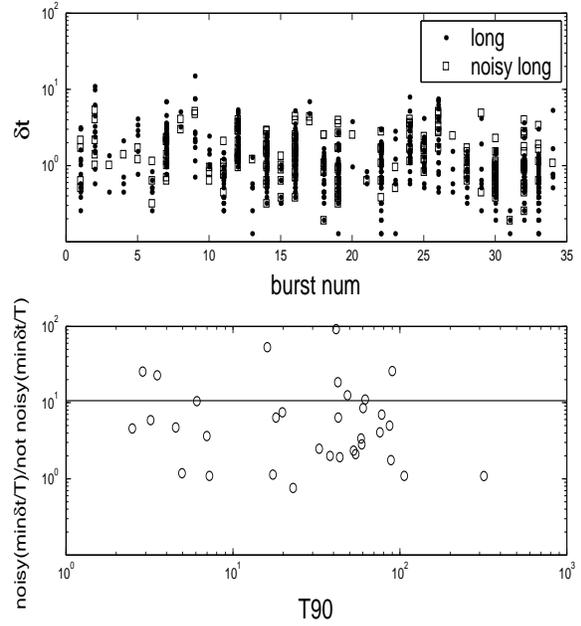,angle=0,height=0.35\textheight,width=0.9\columnwidth}

\caption{\label{noisy VS longs Fig}\textbf{Top (a)} : Pulses
widths in the \textit{'long'} (dots) and \textit{'noisy long'}
\textit{\emph{(squares)}} samples. \textbf{Bottom (b)} : The
ratio between \protect\( \delta t_{min}/T\protect \) in the
\textit{'long'} and \textit{'noisy long'} as a function of BATSE's
\protect\( T_{90}\protect \).}
\end{figure}

Fig.~\ref{noisy VS longs Fig}a represents all the pulses in the
\textit{'long'} and \textit{'noisy long'} samples. The algorithm
retrieves the basic features of the bursts out of the noisy
sample. However, many pulses are 'lost' because of the noise
(only 30\% of the \textit{'long'} pulses are found in the
\textit{'noisy'} sample): (i) Some pulses are too weak to be
distinguished within the amplified noise. (ii) Some pulses merge
with others as the minimum between them are not statistically
significant with the increased noise. The first effect does not
affect the width of the pulses. However, the second one causes
pulse widening. Therefore we expect fewer and wider pulses in the
noisy sample. Both effects are seen clearly in Fig.~\ref{noisy VS
longs Fig}. In the \textit{'noisy'} sample there are 203 pulses
with an average width of 1.62sec while in the original
\textit{'long'} sample there are 695 pulses with an average width
of 1.39sec. The burst duration is affected by the noise as well,
it becomes shorter. This happens when the first or the last
pulses of the burst are lost.

These two effects (pulse widening and shorter burst duration)
tend to increase the value of \( \delta t_{min}/T \).
Fig.~\ref{noisy VS longs Fig}b show that \( \delta t_{min}/T \)
increases by a factor of 10 in average, due to the noise. Thus
the ratio $\delta t_{min} /T$ obtained from a noisy data can be
considered as an upper limit to the real ratio.

\subsection{Attributes  of short bursts pulses\label{shorts result}}

\begin{figure}
\psfig{figure=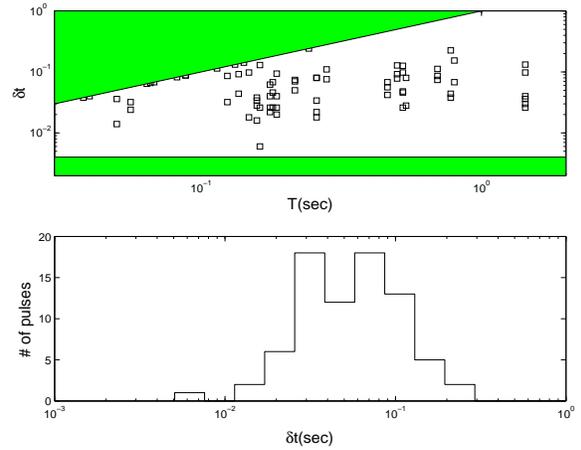,angle=0,height=0.25\textheight,width=0.9\columnwidth}

\caption{\label{shorts pulse width fig}\textbf{Top (a):} The
pulses widths in  short bursts as a function of \protect\(
T\protect \). The gray areas are not allowed because of the
resolution (\protect\( \delta t>4ms\protect \)) or the parameters
definition \protect\( (\delta t \le T\protect \)). \textbf{Bottom
(b)}: A histogram of the pulses widths in short bursts.}
\end{figure}

We have applied the same algorithm to the \textit{`short}' data
sample. The  pulses widths are shown as a function of the bursts
duration in Fig~\ref{shorts pulse width fig} . The gray areas are
not allowed because of the resolution (\( \delta t>4ms \)) or
simply by (\( \delta t\leq T \)). Fig~\ref{shorts pulse width fig}b
depicts the distribution of the pulses width, \( \delta t \). One
can see that  typical values of \( \delta t \) are 50-100 ms with
no significant correlation with T (provided we delete the smooth
single peaked bursts with $\delta t \approx T$).

The pulse width found in this analysis is influenced by several effects. Some
of this effects are due to our algorithm: First, few relatively close pulses
could be seen by the algorithm as a single wide pulse. Second, the width of
two pulses that are not well separated is determined by the minimum between
the pulses. In this case the measured width of both pulses is shorter then their
actual width.

There are also observational effects that influence the pulse
width: First, as indicated in section \ref{noisy vs longs}, the
pulses become wider by a factor of few because of the noise.
Second, the resolution is limited. It is likely that the shortest
pulses in the \textit{'short'} sample are shorter then the best
resolution of our data. Time scales shorter then the data
resolution were already found in short bursts (Scargle, Norris \&
Bonnel 1997).

All the effects described above  except one cause pulse widening.
The exception is when two pulses overlap  leading to a shortening
of the estimated widths of both pulses.   However, this effect
rarely happens in our short bursts analysis. The S/N in this
sample is very low, and the significance level we demand (4\(
\sigma \)) is almost at the signal height (see sec. \ref{shorts}).
Hence, a pulse determined by the algorithm is almost always well
separated (otherwise the minimum between the pulse and its
neighbor would be insignificant). Specifically 61 pulses out of
the 65 pulses  found in  our sample are well separated: the
minimum between pulses, on both sides, is lower than half of the
maximum of the pulse. Thus, pulse widths are almost never
underestimated.  The combination of the other effects causes
pulse widening. Hence our estimate of  \( \delta t_{min}/T \) ,
is only an upper limit.

\begin{figure}
\psfig{figure=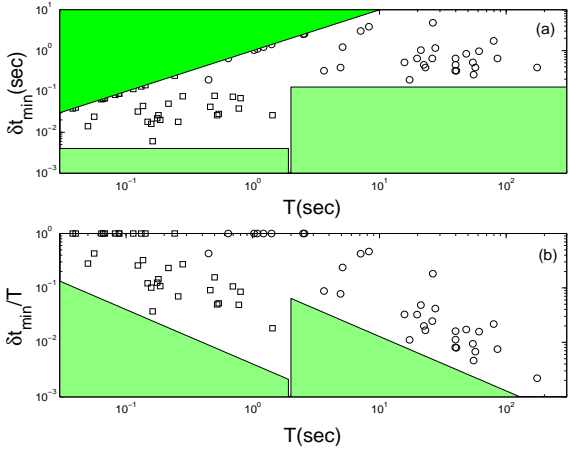,angle=0,height=0.25\textheight,width=0.9\columnwidth}

\caption{\label{dt/T in shorts}
{\bf a)} \( \delta t_{min} \) and {\bf b)} \( \delta t_{min}/T \)
as a function of the total duration of the burst for the {\it 'short'}
and the {\it 'noisy long'} samples. The shaded areas are excluded because of
the data resolution (4ms for shorts and 128ms for noisy longs) or
the parameter definition (\protect\( \delta t_{min}< \le T\protect
\)).}
\end{figure}

Fig.~\ref{dt/T in shorts} shows \( \delta t_{min} \) and \(
\delta t_{min}/T \) for  both groups \textit{'short'} and
\textit{'noisy long'}. In the \textit{'short'} sample the median
\( \delta t_{min}/T \) is 0.25 while 35\% of bursts have \(
\delta t_{min}/T\leq0.1 \) and 35\% of the bursts show a smooth
structure (\( \delta t_{min}/T=1 \)). This result could mislead
us to the conclusion that a significant fraction of the short
bursts have a smooth time profile. But a look at the
\textit{'noisy long'} results show that also in this group more
than 20\% of the bursts are single pulsed, while there were no
such bursts in the original \textit{'long'} sample\textit{.}
Naturally, the bursts that loose the fine structure because of
the noise are bursts with fewer original pulses. It is clear that
short bursts have less pulses then long ones and therefore they
are more ``vulnerable{}'' to this effect. Hence we conclude that
at least one third of short bursts are highly variable (\( \delta
t_{min}/T\ll 1 \)) and it is very likely that another third
(those bursts with \( 0.1< \delta t_{min}/T < 1 \))is variable as
well. We cannot tell whether the smooth structure (\( \delta
t_{min}/T = 1 \)) seen in a third of the short bursts  is
intrinsic or whether it arises due to the noise.

\begin{figure}
\psfig{figure=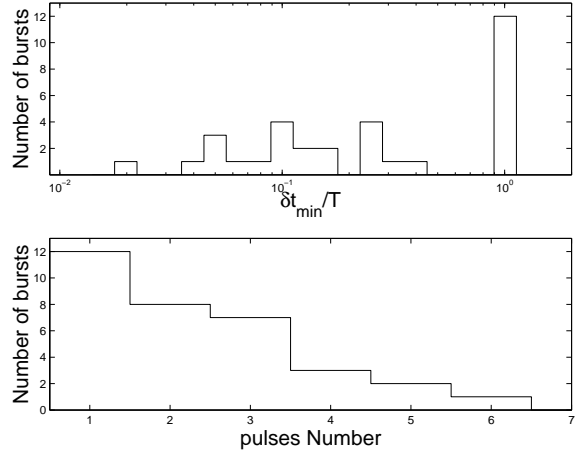,angle=0,height=0.25\textheight,width=0.9\columnwidth}

\caption{\textbf{Top}: The histogram of  $\delta t_{min}/T$ in short bursts.
\textbf{Bottom:} The histogram of the pulses number within a burst in the {\it `short'} sample.}
\end{figure}

While we are mostly interested in this analysis in the pulse
width, it is worth mentioning that the amplitude of the
variations is large. In most (61 out of 65) cases, when there is
no overlap of nearby pulses the number of counts drops to less
than half of the maximal counts. Thus the pulse we describe are
not only statistically significant, they also correspond to a
significant (factor of 2) variation in the output of the source.

\subsection{High resolution analysis of long bursts\label{long start result}}

\begin{figure}
\psfig{figure=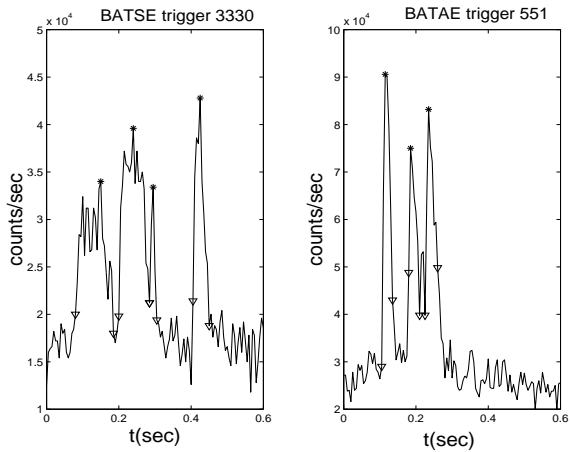,angle=0,height=0.25\textheight,width=0.9\columnwidth}

\caption{\label{burst3330 vs burst551}\textbf{Left}) The
beginning of BATSE trigger 3330 (a long bright burst with
\protect\( T_{90}=62sec\protect \) ). \textbf{Right}) The whole
light curve of BATSE trigger 551 (a bright short burst with
\protect\( T_{90}=0.25sec\protect \)). The peaks found by our
algorithm marked by stars. The triangles mark the pulses width.
The figure demonstrates  the similarity of short time scale
structure  in these bursts  (at a 5 msec resolution).}
\end{figure}

We compare the time profile of the first seconds of 15 long
bursts with the time profiles of 15 short bursts. Fig.
\ref{burst3330 vs burst551} shows the light curves of a short
burst and the first second of a long burst. The time scales of
both bursts are quite similar. It is difficult to decide, on the
base of these light curves alone, which one belongs to a short
burst and which one is a fraction of a long one.

\begin{figure}
\psfig{figure=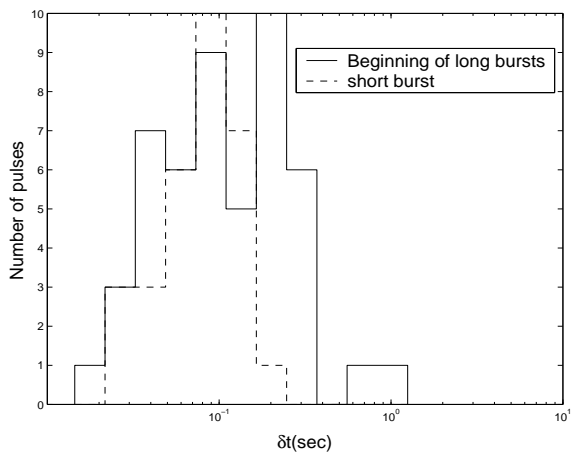,angle=0,height=0.25\textheight,width=0.9\columnwidth}

\caption{\label{start longs fig} The histograms of pulses widths
in the initial 1-2 seconds of long bursts (\textbf{Smooth line})
and  pulses widths in short bursts (\textbf{Dashed line}). Both
samples have a 5msec time bins.}
\end{figure}

Fig.~\ref{start longs fig} shows the pulse widths histograms of
the beginning of long bursts and of the short bursts. The time
scales in both samples are quite similar in the range of
10-200ms. The long bursts have additional pulses in the range of
0.2-1sec. Long bursts contain, of course, longer pulses, but in
the sample we considered we demanded that the counts would fall
back to the background level within the TTE data. In this way we
have limited the pulses width of the long bursts. Both histograms
begin at 10-20ms, which is at the limit of the pulse width
resolution (10ms). It is likely that both samples contain shorter
time scales that cannot be resolved.

The counts variations within the pulses observed in the high
resolution long bursts is high. In 41 out of 46 pulses, the
photon counts drops to less than half of the maximal counts of
the pulse on both sides. Therefore, like in short bursts, the
statistically significant variations in the bursts reflects also a
significant variation in the output of the source.

Walker, Schaefer \& Fenimore (2000) performed a similar analysis
of TTE data of 14 long bursts. They found only one long burst
with very short time scales. The difference between the results
arises from the different samples considered. Walker et al.
(2000) considered the bursts with maximal total photon counts
within the TTE burst record.  We demanded that the counts will
return close to the background level within the TTE record.
Walker et al. (2000) criterion  favors  bursts that are active
during the whole TTE record and therefore most of the bursts in
their sample are dominated by a long and bright pulse.

\begin{figure}
\psfig{figure=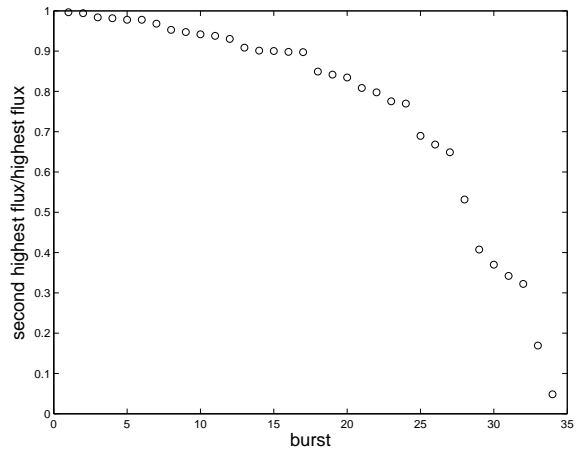,angle=0,height=0.25\textheight,width=0.9\columnwidth}

\caption{\label{flux ratio} The ratio of the fluxes of the second
brightest and the brightest peaks in long bursts.}
\end{figure}

This similarity between the short bursts and the fraction of long
bursts raises the question whether it is possible that short
bursts are actually only a small fraction, which is above the
background noise, of long bursts. We have already seen that the
noise cause us to loose pulses. Is it possible that a long burst
with a single dominant, very intense pulse, (or a group of very
close and intense pulses) will loose all its structure, apart for
this intense pulse, due to noise and become a short burst. Fig.
\ref{flux ratio} rules out this option. It depicts the counts
ratio between the most intense pulse and the second most intense
pulse within long bursts. The graph shows that in 32 out of 34
bursts the second highest peak is more then third  of the most
intense peak and in 27 bursts the second peak is more than two
thirds of the first. The noise cannot cause one pulse to
disappear without the other. As these two peaks  are usually
separated by more than two seconds, the noise cannot convert a
significant fraction of long bursts to short ones. Clearly the
noise  has some effect on the duration histogram and in  a few
cases the added noise converted long bursts into a short ones, but
it certainly cannot produce the observed bimodality.

\section{Discussion}

We have shown that most short bursts as well as most long bursts
are multi-peaked and highly variable with \( \delta t_{min}/T\ll 1
\). These statistically significant variations involve generally a
change of more than a factor of two in the count rate. 
30\% of the short bursts have a single pulse for which \( \delta
t_{min} \) is the same as the observed duration - \( \delta
t_{min}\approx T \). These are smooth bursts. However, a
comparison with the \textit{'noisy long'} sample, shows that this
might be an artifact of the low S/N in the short bursts sample.
We also find significant variability on very short time scale in
long GRBs. This  reduces the values of the variability parameter
\( \delta t_{min}/ T \) to $10^{-3}-10^{-4}$ in long bursts.

External shocks (at least simple models of external shocks) cannot
produce variable bursts (Sari \& Piran, 1997; Fenimore, Madras \&
Nayakshin, 1996). Our result suggests that most short bursts are
produced via internal shocks. 30\% of the short bursts are
smooth.  We cannot rule out the possibility that these bursts are
produced by external shock. The observed very short time scales
in long GRBs strengthen further the argument in favor of internal
shocks in these bursts and requires more contrived and fine tuned
external shocks models for variability.

Kobayashi, Piran \& Sari (1997) have shown that internal shocks
cannot convert all the relativistic kinetic energy  to gamma-rays.
The remaining kinetic energy is dissipated latter when the
relativistic ejecta is slowed down by the surrounding medium. The
resulting external shocks produce the afterglow. An essential
feature of this internal-external shocks scenario is that the
afterglow is not a direct extrapolation of the initial
$\gamma$-ray emission. This model suggests (Sari, 1997) for long
bursts an overlap between the GRB and initial phase of afterglow.
This have been indeed observed in several cases as a build up of
a softer component during the GRB and a corresponding transition
in the spectrum of long GRBs to a soft x-ray dominated stage
towards the end of the burst.

For short GRBs the internal-external scenario suggests that the
afterglow will begin few dozen seconds after the end of a short
burst that was produced by internal shocks (Sari, 1997). It also
suggest that this afterglow will not be a direct extrapolation of
the GRB (as the two are produced by different mechanisms). Thus
we predict that for most short bursts there will be a gap between
the burst and the beginning of the afterglow.  We have already
remarked that at this stage one cannot tell whether the remaining
30\% smooth short bursts are produced by external or by internal
shocks. This could be tested in the distant future with better
data. However, there are clear predictions concerning the
afterglow that could distinguish between the two possibilities.
If these short GRBs are produced by internal shocks then we
expect a clear gap between the end  of the GRB and the onset of
the x-ray afterglow. If on the other hand these short bursts are
produced by external shocks then we predict that the afterglow
would continue immediately with no interpretation after the GRB
and its features  would be a direct extrapolation of the
properties of the GRB.

\section*{Appendix -The algorithm}

Our algorithm finds the peaks of the bursts. Each peak corresponds to a single
pulse; a pulse is the basic event of the light curve. The algorithm is based
on the algorithm suggested by Li \& Fenimore (1996). Li \& Fenimore define a
time bin \( t_{p} \) (with count \( C_{p} \)) as a peak if there are two time
bins \( t_{1}<t_{p}<t_{2} \) (with counts \( C_{1},C_{2} \) respectively)
which satisfies a) \( C_{p}-C_{1,2}>N_{var}\sqrt{C_{p}} \) and b) \( C_{p} \)
is the maximal count between \( t_{1} \) and \( t_{2} \). \( N_{var} \) is
a parameter that determines the significance of the peak.

There are two problems with this algorithm. First, this algorithm analyzes only
data in a single time resolution (fixed time bin size). Therefore the algorithm
looses long and faint pulses. A peak that does not satisfy the criterion described
above in the raw data resolution could satisfy the criterion if the data resolution
is lower (longer time-bins). This algorithm would miss such a pulse. Second,
\( N_{var} \) determines the trade-off between sensitivity and false peaks
identification rate. When \( N_{var} \) is low the algorithm finds false peaks
as a result of the Poisson noise. When \( N_{var} \) is high the algorithm
misses real peaks. Finding false peaks is a severe problem during long periods
of constant level Poisson noise, like the background (as will be explained shortly).
Long bursts contain such periods (periods of only background noise). These periods
are called quiescent times. In order to avoid false peaks in long bursts
\( \textrm{N}_{\textrm{var }} \)must
be large(\( \geq 7 \)), which means an insensitive algorithm. Short bursts
contain less quiescent times, and of course shorter ones. But, short bursts
contain much less pulses then long burst (three to four compared to an average
of more than thirty) and much smaller S/N. Finding even one false peak could
change the features of the burst drastically. Too insensitive algorithm could
loose all the burst structure. In order to avoid false peaks in short bursts
\( \textrm{N}_{\textrm{var }} \)must be at least as large as 5. As described
in section \ref{shorts}, the S/N in some of the bright short bursts is smaller
than 5. Such \( N_{var} \) will prevent the algorithm from finding even one
peak in these bursts.

We solved the first problem by analyzing the data in different resolutions.
The results of the algorithm in different resolutions are merged into a single
sample of peaks. We solved the second problem by restricting the search for
peaks only to `Active Periods'. Active periods are periods with counts that
correspond to source activity (we will define it later on).

There are few advantages for analyzing only active periods. The main one is
that a lower \( N_{var} \) can be used during these periods with smaller risk
of finding false peaks. The risk of finding false peaks due to a Poisson noise
depends on the time scale in which the original signal (i.e. without the noise)
changes its count rate in the same order as the Poisson noise level. If the
signal is constant (no real pulses), then this time scale is as long as the
signal. If the constant signal is longer, it contains more time bins with the
same level of Poisson noise counts. Hence, there is a larger chance of finding
within these time-bins three time-bins, \( t_{p} \), \( t_{1} \) and \( t_{2} \),
that satisfy the criterion in Li \& Fenimore algorithm. Then \( t_{p} \) would
become a false peak. On the other hand, if the signal is changing monotonically
then the length of the signal is irrelevant. \( t_{p} \), \( t_{1} \) and
\( t_{2} \) must be within the period in which the signal changes at the same
order as the Poisson noise level; there is no chance of finding \( t_{2} \)
with \( C_{2} \) significantly below \( C_{p} \) (if the signal is rising)
out of this period. During the active periods the signal is changing rapidly
(usually on time scales of seconds or less), and the Poisson noise is superimposed
on steep slopes. In this case a false peak could only be found during the period
in which the signal didn't change compared to the noise level. There are much
less time bins during this period and hence there are much less chance of finding
false peaks.

The second advantage is that when an active period is found we almost certain
that it is a part of the burst. This is important since one false peak in the
`wrong' place (for example hundred of seconds after the burst ended) can change
the burst properties drastically. By analyzing only active periods we can use
smaller \( N_{var} \) (=4) and get a more sensitive and accurate algorithm.

\begin{figure}
\psfig{figure=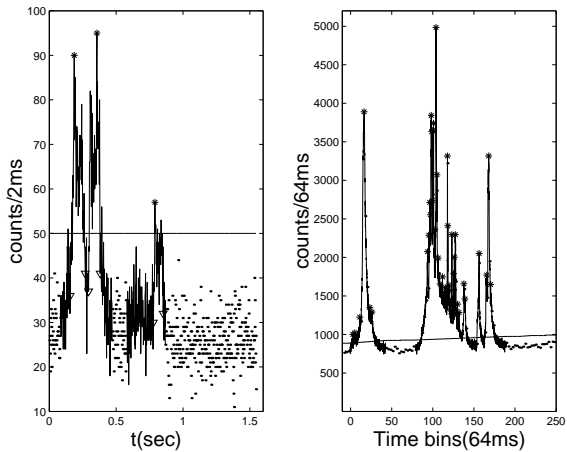,angle=0,height=0.25\textheight,width=0.9\columnwidth}

\caption{\label{algorithm exsample fig} {\bf Left)} Time profile of BATSE trigger 2952 (bright
short burst). {\bf Right)} Time profile of BATSE trigger 2156 (bright long burst).
The solid line marks the active periods. The horizontal line marks the activity
level. All the time bins with counts above the activity level are `active bins'.
The peaks are marked by '{*}'.}
\end{figure}

Our algorithm works in several steps. First, it determines the background level
of the signal (as a function of time). Then it finds an activity level, demanding
a probability of 0.9 (per burst) that all the time bins with counts above this
level (called `active bins') correspond to source activity and not of the background
Poisson noise (we demand that on every ten bursts there is, on average, a single
false `active bin'). The activity level depends on the background and its value
is between 4\( \sigma  \) to 5\( \sigma  \) above the background. From each
active bin we search to the right and to the left until the count level drops
to the background level on both sides. We call all these bins together an active
period (from the time bin that the counts are above the background until the
time bin that the counts reaches the background level again). In most cases
a single active period includes many active bins and a burst may contain more
then a single active period (see Fig~\ref{algorithm exsample fig}). Note that
if the algorithm misses an active period in one resolution, it can still find
it in a different (lower) resolution, in which the noise level is lower.

Once the active periods of a given burst have been determined we apply the Li
\& Fenimore algorithm to the active periods (using Nvar=4) and determine the
peaks. We repeat this procedure (finding the active periods and the corresponding
peaks), several times for different time resolution. To obtain lower resolution
data we convolve the original signal (in the basic time bins) with a Gaussian,
whose width determines the resolution. Finally, after finding the peaks in different
resolutions we merge those samples of peaks to a single sample (requiring that
a peak must appear in at least two different resolutions). The merge is done
by merging the highest resolution sample with the second highest one and then
taking this merged sample and merging it with the third highest resolution sample
and so on. On different resolutions the same peak could be found on different
time bins. In each case two peaks on different resolutions are considered as
a single one if the peak in one resolution falls between \( t_{1} \) and \( t_{2} \)
of the peak in the other resolution.

Each peak corresponds, of course, to a pulse. The pulse width (\( \delta t \))
is defined by two points (on each side of the peak) that are higher than the
background by 1/4 of the peaks height or by the minimum between two neighboring
peaks (if the latter is higher). The duration (\( T \)) of the burst is the
time elapsed from the beginning of the first pulse till the end of the last
pulse (so in single pulsed burst \( T=\delta t \)).

\section*{Acknowledgments}

This research was supported by US-Israel BSF grant.


\begin{thebibliography}{10}
\bibitem{1a}Beloborodov A. M., Stern B. E., Svensson
R., 2000, ApJ, 535, 158
\bibitem{2}Cline D. B., Matthey C.,
Otwinowski S., 1999, ApJ, 527, 827
\bibitem{1}Kobayashi S., Piran T. Sari R., 1997, ApJ, 490, 92
\bibitem{2}Lee A., Bloom E., Scargle J., 1995, in Kouveliotou
C., Briggs M. S.,  Fishman G.J., Eds., AIP Conf. Proc. 3rd Huntsville
Symposium,  Gamma-Ray Bursts, p.47 (New York: AIP)
\bibitem{3}Lee A., Bloom E. D., Petrosian V., 2000, ApJS, 131, 21
\bibitem{4}Li H., Fenimore E. E., 1996, ApJ, 469, L115
\bibitem{5} Fenimore E. E., Madras C. D., Nayakshin
S., 1996, ApJ, 473, 998
\bibitem{6}Norris J. P., Nemiroff R. J., Bonnell J. T., Scargle J. D.,
  Kouveliotou C., Paciesas W. S., Meegan C. A., Fishman G. J.,
  1996, ApJ, 459, 393
\bibitem{7}Norris J. P., 1995, in Kouveliotou
C., Briggs M. S.,  Fishman G.J., Eds., AIP Conf. Proc. 3rd Huntsville
Symposium,  Gamma-Ray Bursts, p.13 (New York: AIP)
\bibitem{8}Sari R.,1997, ApJ, 489, L37
\bibitem{9}Sari R., Piran T., 1997, ApJ, 485, 270
\bibitem{10}Scargle J. D., 1998, ApJ, 504, 405
\bibitem{11}Scargle J. D., Norris J., Bonnel J., 1997 in Meegan C.,
Preece R., Koshut T., Eds., AIP Conf. Proc. 4th Huntsville
Symposium,  Gamma-Ray Bursts, p.181 (New York: AIP)
\bibitem{12}Walker K. C., Schaefer B. E. , Fenimore, E. E., 2000, ApJ, 537, 264 \end{thebibliography}
\end{document}